\documentstyle[a4wide]{article}

\newcommand {\xivec} {\mbox{\boldmath $\xi$\unboldmath} }
\newcommand {\Bvec} { {\rm \bf B}}
\newcommand {\Jvec} { {\rm \bf J}}
\newcommand {\erf} {{\rm erf}}

\input{psfig}

\begin{document}

\date{}

\title{\bf Phase transitions in soft--committee machines} 

\author { M.Biehl, E. Schl\"osser,  and M. Ahr \\
            Institut f\"ur Theoretische Physik,
            Julius--Maximilians--Universit\"at W\"urzburg\\
            Am Hubland, D--97074 W\"urzburg, Germany}

\iffalse
\rec{}{}

 \fi

%
\maketitle

\begin{abstract}
Equilibrium statistical physics is applied to layered neural networks 
with differentiable activation functions.
A first analysis of off--line learning in soft--committee machines with 
 a finite number $(K)$ of hidden units learning a perfectly matching rule is performed.
Our results are exact in the limit of high training temperatures $(\beta \to 0)$. 
For $K=2$ we find a second order phase transition from  unspecialized to 
specialized student configurations at a critical size $P$ of the training set, 
whereas for $K\geq 3$ the transition is first order. 
Monte Carlo simulations indicate that our 
results are also valid for moderately low temperatures qualitatively.
The limit $K\to\infty$ can be performed analytically, the transition occurs after presenting  
on the order of $ N K/\beta$ examples. However, an unspecialized metastable state  
persists up to $P \propto N K^2/\beta$.  
\end{abstract}

\ \\

Statistical physics provides tools for the investigation of
learning processes in adaptive systems such as 
feedforward neural networks \cite{Hertz}. 
The by now  standard analysis of off--line or batch learning 
from a fixed set of example data is based on the interpretation of  training  
as a stochastic process which leads to a properly defined thermal equilibrium.
It has been applied with great success to simple networks like single layer
preceptrons or specific multilayer architectures with binary threshold units.
See e.g.\  \cite{KinzelOpper,SST,review} 
for reviews and \cite{phase} for
a discussion focussed on phase transitions in neural networks. 

In a somewhat different framework, the theory of on--line learning,
it is  assumed that a temporal sequence of independent examples is 
provided by the environment. For large networks, the inherently stochastic 
learning dynamics is described exactly  by a set of 
deterministic differential equations, see e.g.\ \cite{Cambridge} for an up
 to date overview.  This approach has made possible the recent  progress 
with respect to layered networks with continuous node activations,
see e.g.\ [6-10]
%\cite{BiehlSchwarze,SaadSolla,BRW,local,global} 
and references therein.
Such systems are relevant for applications as they can 
implement non--trivial regression schemes and practical training
algorithms are available \cite{Hertz,Chauvin}.

In this Letter we present an  analysis of off--line learning in 
two--layered architectures with differentiable transfer functions 
by means of equilibrium statistical physics. 
The considered model exhibits phase transitions in the learning process, 
i.e.\ a discontinuous dependence of the student performance on the number 
of examples.  These transitions are the counterparts of quasi--stationary 
plateau states observed in on--line dynamics \cite{SaadSolla,BRW} and are
due to the same inherent symmetries.  
The hidden unit specialization studied here is different from the sudden 
achievement of perfect generalization  described in \cite{boesetal} for single,
continuous nodes.  

Specifically, we will investigate in the following the learning of a rule in a
fully connected two--layered neural network with total output 
\begin{equation} \label{studentoutput} \textstyle
 \sigma (\xivec) \, = \, \frac{1}{\sqrt{K}} \, \sum_{j=1}^{K} \, g(x_j) 
  \mbox{~~~where~~} x_j = \frac{1}{\sqrt{N}} \, \Jvec^{(j)} \cdot \xivec
 \end{equation}
 upon presentation of an $N$--dimensional input vector $\xivec$.
 Given the hidden unit activation $g(x)$, the  adaptive weights 
 $\Jvec^{(i)} \in I\!\!R^N$ define the input--output relation. 
 The term  {\it soft--committee machine \/} has been coined for this type of 
 network \cite{BiehlSchwarze,SaadSolla},
 as it can be interpreted as a continuous version of the thoroughly 
 studied ({\sl hard\/}) committee of binary hidden units (see [14-17]
%\cite{oh,HolmJohn,Holmnur,oh} 
 and references therein). 
 Here, the weights of the linear hidden--to--output relation are fixed
 to the  particular value $1/\sqrt{K}$ which  
 differs from the usual scaling considered in the analysis of on--line learning
 in soft--committees \cite{SaadSolla,BRW}.

 Off--line learning in networks of type (\ref{studentoutput}) has been studied 
 before in the limit $K \to \infty$ by Kang {\sl et al.} \cite{oh}. Here we will
 focus on finite $K$, but include a discussion of large machines for completeness.
 Note that the specific form of Eq.\ (\ref{studentoutput}) yields outputs of finite 
 magnitude in the limit $K \to \infty$.
 Most frequently $g(x)$ is taken to be a sigmoidal function of its argument, e.g.\ 
 the hyperbolic tangent.  We choose the similiar but more convenient function 
 $ g(x) = \erf(x/\sqrt{2}) $ which simplifies the mathematical treatment to a great extent 
 yet should not alter the basic features of the model  otherwise 
 \cite{BiehlSchwarze,SaadSolla}.

 In this Letter, we restrict our analysis to scenarios in which 
 the unknown rule $\tau(\xivec)$ 
 can be parametrized through a teacher network 
 of perfectly matching architecture and size $(K)$ with weight vectors
 $\Bvec^{(j)}$.  
 Further, we assume that the  $\Bvec^{(j)}$ are normalized vectors  
 of length $\sqrt{N}$  with i.i.d.\ random components 
 and  accordingly impose a normalization $ (\Jvec^{(j)})^2 = N $ on the student
 vectors.  Since the continuous student output, Eq.\ (\ref{studentoutput}),
 depends explicitely on the length of the weight vectors,  
 this latter constraint corresponds to significant
 {\it a priori\/} knowledge of the rule's structure. We will discuss later how
 this restriction could be relaxed.

 The  off-line  training process is based on  
 a fixed set of examples $ \left\{ \xivec^{\mu}, \tau(\xivec^{\mu})
  \right\} $~~$(\mu =1,2,\ldots, P)$ and  is guided by the minimization of
 the cost function or training error  
 \begin{equation} \label{trainerror} \textstyle
 \varepsilon_t \, = \, \frac{1}{P}  \sum_{\mu=1}^P \,
  \frac{1}{2} \left[ \sigma(\xivec^\mu) - \tau(\xivec^\mu)  \right]^2.
 \end{equation} 
 Thus, learning is formulated as an optimization problem.
 In networks with differentiable  activation functions the 
 (approximate)  solution  could be found by use of gradient descent
 or similiar  methods.  The prominent {\it backpropagation of error\/}, 
 for example, is widely used in practice \cite{Hertz,Chauvin}.
 Often, the actual global minimum of (\ref{trainerror}) is not identified by
 such algorithms, even if the rule is perfectly learnable.
 Practical learning prescriptions  can be trapped
 in local minima of the training error or  stop as soon as a satisfactory
 performance is achieved.

 After training the  quadratic error measure can also be utilized to quantify
 the success of learning in terms of the so--called generalization error 
 \begin{equation} \label{gendef} \textstyle
  \varepsilon_g  \, = \, \frac{1}{2} \left\langle  \left[\sigma(\xivec) - \tau(\xivec)\right]^2
  \right\rangle.  \end{equation}
 Here $ \left\langle \ldots \right\rangle $  denotes an empirical  average over a 
 test set of (new) examples  or  over the distribution of random inputs, which is assumed to be
 known in the model. 
 Throughout the following we take the components of all inputs $\xivec$ to be i.i.d.\ random
 variables with zero mean and unit variance. 
 In the thermodynamic limit $N\to\infty$
 the quantities $x_j = \Jvec^{(j)}\!\cdot\!\xivec /\sqrt{N}$
 and $ y_j = \Bvec^{(j)}\!\cdot\!\xivec /\sqrt{N}$ 
 become  zero mean correlated Gaussian variables by means of the central limit theorem.
 Their joint density is fully characterized by the covariances
 \begin{equation} \textstyle
   \left\langle x_i  x_j \right\rangle = \frac{1}{N} \Jvec^{(i)}\!\cdot\!\Jvec^{(j)} = \,
     Q_{ij},  \,\,
   \left\langle x_i  y_j \right\rangle =  \frac{1}{N} \Jvec^{(i)}\!\cdot\!\Bvec^{(j)} = \,
     R_{ij},  \mbox{~~and~~} 
   \left\langle y_i  y_j \right\rangle =  \frac{1}{N} \Bvec^{(i)}\!\cdot\!\Bvec^{(j)} = \,
     \delta_{ij} 
   \label{covariances}
\end{equation}
where all diagonal $Q_{ii} =1$.
Hence, the average in Eq.\ (\ref{gendef}) reduces to a $2K$--dimensional Gaussian
integral which can be performed analytically  for the specific activation function 
$g(x) = \erf(x/\sqrt{2})$.  The result
depends only on the {\it order parameters\/} $\left\{R_{ij},Q_{ij}\right\}$:
\begin{equation} \label{generrorgeneral} 
\varepsilon_g  \left(\left\{Q_{ij},R_{ij}\right\}\right) \,= \, \frac{1}{6} + \, \frac{1}{\pi} 
  \sum_{i,j=1}^K \, \left[ \arcsin\left(\frac{Q_{ij}}{2}\right)  - 2 
\arcsin\left(\frac{R_{ij}}{2}\right)  \right]
\end{equation}
and 
was derived in \cite{SaadSolla} for arbitrary network sizes and more general scenarios.

Following the standard statistical physics approach to off--line learning we
 consider a Gibbs ensemble of networks which is characterized by the 
 partition function
 \begin{equation} \label{partition} \textstyle
  Z \, = \, \int d\mu (\{ \Jvec^{(i)} \}) \,
  \exp \left[ -\beta \, P \, \varepsilon_t   \right].
 \end{equation}
 The extensive  energy $ P \, \varepsilon_t $ is defined in Eq. (2) and the 
 measure $ d\mu $ limits the integration to the  region in weight space
 where all $(\Jvec^{(i)})^2 = N$.  
 The formal temperature $1/\beta$ controls the thermal average  of the energy in 
 equilibrium. Equivalently, it fixes the amount of noise which is 
 present in a corresponding stochastic training process.  

Typical properties of the equilibrium state can be calculated from the associated
free energy on average over the quenched randomness contained in
the training data $ I\!\!D = \left\{ \xivec^{\mu} , \tau (\xivec^\mu) \right\}$ (denoted
as $\left\langle \ldots \right\rangle_{I\!\!D}$).
The evaluation of the quenched free energy $ - \left\langle \ln Z \right\rangle_{I\!\!D} / \beta $ 
 requires, in general, the application of
the replica method. It becomes rather involved in models of the
complexity considered here, see e.g.\ 
 [15-17] for a treatment of networks with binary units.
 In order to obtain first results for the off--line training of soft--committee machines  
 we resort to the simplifying limit of high temperatures. 
 This strategy  has proven useful for gaining first insights in a variety of models,
 see  \cite{SST,review,SOK} for details and example applications.

 In the limit $\beta \to 0$, we can  replace $ \left\langle  \ln Z \right\rangle_{I\!\!D} $
 with $\ln \left\langle Z \right\rangle_{I\!\!D}$, i.e.\ the annealed approximation,  which
 circumvents the replica formalism, becomes exact. Further, the average over the training
 data factorizes with respect to the example inputs and one obtains 
 \begin{equation} \label{freeenergy}
  - \left\langle \ln Z \right\rangle_{I\!\!D} / N
  \, = \beta f (\{Q_{ij},R_{ij}\}) \, = \, \, 
   \left(\beta P / N \right)  \,\,\,  \varepsilon_g (\{Q_{ij},R_{ij}\})
    - s \left(\left\{Q_{ij},R_{ij}\right\}\right)   
 \end{equation}
 where the r.h.s.\ is to be minimized with respect to the order parameters.
  Non--trivial results can only be expected if the effective temperature
  $ (\beta P / N) $ is of order $1$, i.e.\ the high training temperature has
  to be compensated for by a large number of examples $ P \propto N/\beta$.   
  Consequently, training energy $\varepsilon_t$  and generalization error 
 $\varepsilon_g$ coincide in this limit. 
 Eq.\  (\ref{freeenergy}) contains the  entropy term 
 \begin{equation} \label{entdef}  \textstyle
  s (\{Q_{ij},R_{ij}\})\, =  (1/N) \, \ln  \int \prod_{j}  d^N {\rm J}^{(j)} \,
    \prod_{i,j} \, \left(  \delta ( \Jvec^{(i)}\!\cdot\!\Bvec^{(j)}\!-\!NR_{ij} ) 
    \delta (\Jvec^{(i)}\!\cdot\!\Jvec^{(j)}\!-\!NQ_{ij} ) \right)
 \iffalse
  s (\{Q_{ij},R_{ij}\})\, =  \frac{1}{N}  \ln  \int \prod_{j=1}^{K}  d^N {\rm J}^{(j)} \,
    \prod_{i,j=1}^{K} \, \left(  \delta ( \Jvec^{(i)}\!\cdot\!\Bvec^{(j)}\!-\!NR_{ij} ) 
    \delta (\Jvec^{(i)}\!\cdot\!\Jvec^{(j)}\!-\!NQ_{ij} ) \right)
\fi
 \end{equation}
 which can be evaluated by means of a saddle point integration 
 after rewriting the $\delta$--functions in their integral representation.  One obtains
 $s (\{Q_{ij},R_{ij}\}) \, = \,  \ln[\det {\cal C}]/2  + {\rm const,} $ 
 where ${\cal C}$ is the $(2K\times2K)$--matrix of all cross-- and self--overlaps  of the
 vectors $\{ \Jvec^{(j)}, \Bvec^{(j)} \}$. The constant term is independent
 of the  order parameters and therefore irrelevant.

 In order to proceed with the analysis,  we assume  
 that the  equilibrium student configuration is symmetric with
 respect to the hidden--units:
 \begin{equation} \label{sitesymmetry}  \textstyle
  R_{ij} = \, R  \, \delta_{ij} \, + \, S \, (1-\delta_{ij}) \mbox{~~~~and~~}
  Q_{ij} = \,  \delta_{ij} \, + \,  C \, (1-\delta_{ij}).
  \iffalse
   R_{ij} = \left\{  \begin{array}{ll}  R & \mbox{for~~} i = j \\
                                        S & \mbox{for~~} i \neq j \\ \end{array} \right.
   \qquad \mbox{and} \qquad
    Q_{ij} = \left\{  \begin{array}{ll}  1 & \mbox{for~~} i = j \\
                                        C & \mbox{for~~} i \neq j \\ \end{array} \right..
  \fi
 \end{equation}
 This assumption reflects the symmetry of the rule, yet allows for
 the {\sl specialization\/} of student nodes: for $R > S$  each of 
 them has achieved a larger overlap with exactly one of the teacher vectors. 
 In the limiting case $ R\!=\!1, S\!=\!C\!=\!0 $
 the student is identical with the teacher  and generalizes perfectly $(\varepsilon_g =0)$. 
 Now entropy (apart from irrelevant constants) and generalization error read 
 \begin{eqnarray} \label{freesym}
  s &=& \frac{1}{2} \ln \left[1+\!(K\!-\!1) C \!- \!\left((R\!-\!S) \!+ \!KS\right)^2 \right] 
       + \frac{K-1}{2} \ln \left[ 1\!-\!C\!-\!(R\!-\!S)^2 \right]  \\ 
 \varepsilon_g  & = & \frac{1}{3} + \frac{K-1}{\pi} \left[ \arcsin \left(\frac{C}{2}\right)
  - 2\arcsin\left(\frac{S}{2}\right) \right] -
\frac{2}{\pi} \arcsin\left(\frac{R}{2}\right).  \label{gensym} 
 \end{eqnarray}

  Defining $ \alpha = \beta P / (NK) $, the rescaled number of examples per student weight,
  we  have minimized  $f = \alpha (K \varepsilon_g) - s$ with respect to the three 
  order parameters $R,C,S$ numerically for different network sizes. 

  For $K=2$ we find that the equilibrium configuration is characterized by 
  $ R = S$ for  $\alpha < \alpha_c^{(2)} \approx 23.7$,
  whereas above this critical value the only solution obeys $ \Delta = |R - S| \neq 0$. 
  The system undergoes a second order phase transition and the specialization  $\Delta$ 
  increases  close to the critical point like $(\alpha- \alpha_c^{(2)})^x$ where $x=0.5$.
  Figure 1 shows that  this spontaneous symmetry breaking translates into a kink in the 
  learning curve $\varepsilon_g (\alpha)$. 

\begin{figure}[t]
\begin{center}
\setlength{\unitlength}{0.8pt}
\begin{picture}(325,150)(0,0)
 \put(0,0){\makebox(325,150)
          {\includegraphics{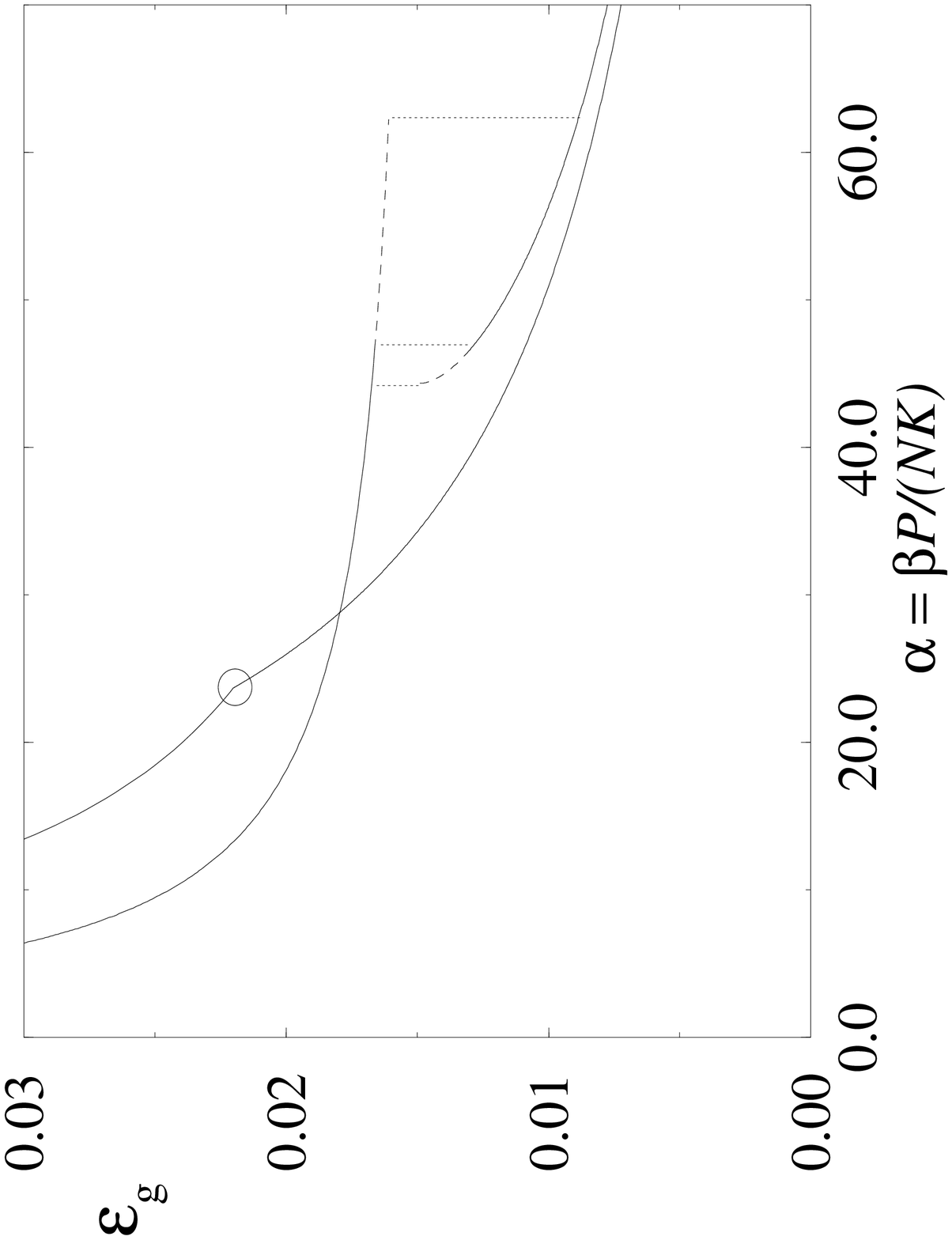}}}
  \put(0,0){\makebox(-140,-35){\bf a)}}
% \put(0,0){\makebox(-170,280){\large $\varepsilon_g$ }}
% \put(0,0){\makebox(100,-30){ $\alpha = P/(NK)$ }}
 \put(0,0){\makebox(325,150)
          {\includegraphics{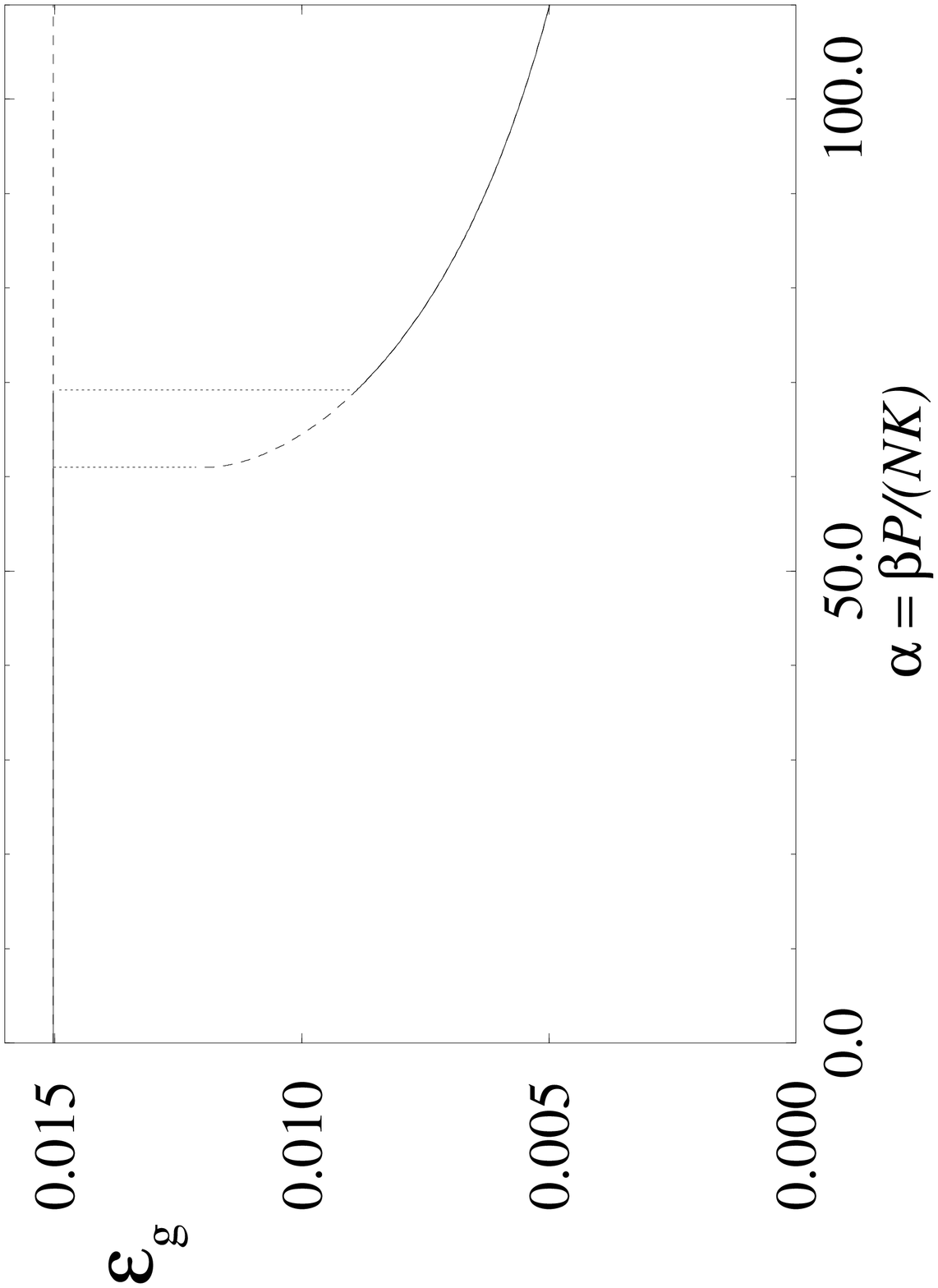}}}
 \put(0,0){\makebox(400,-35){\bf b)}}
 \put(174,105){\makebox(80,40)
      {\includegraphics{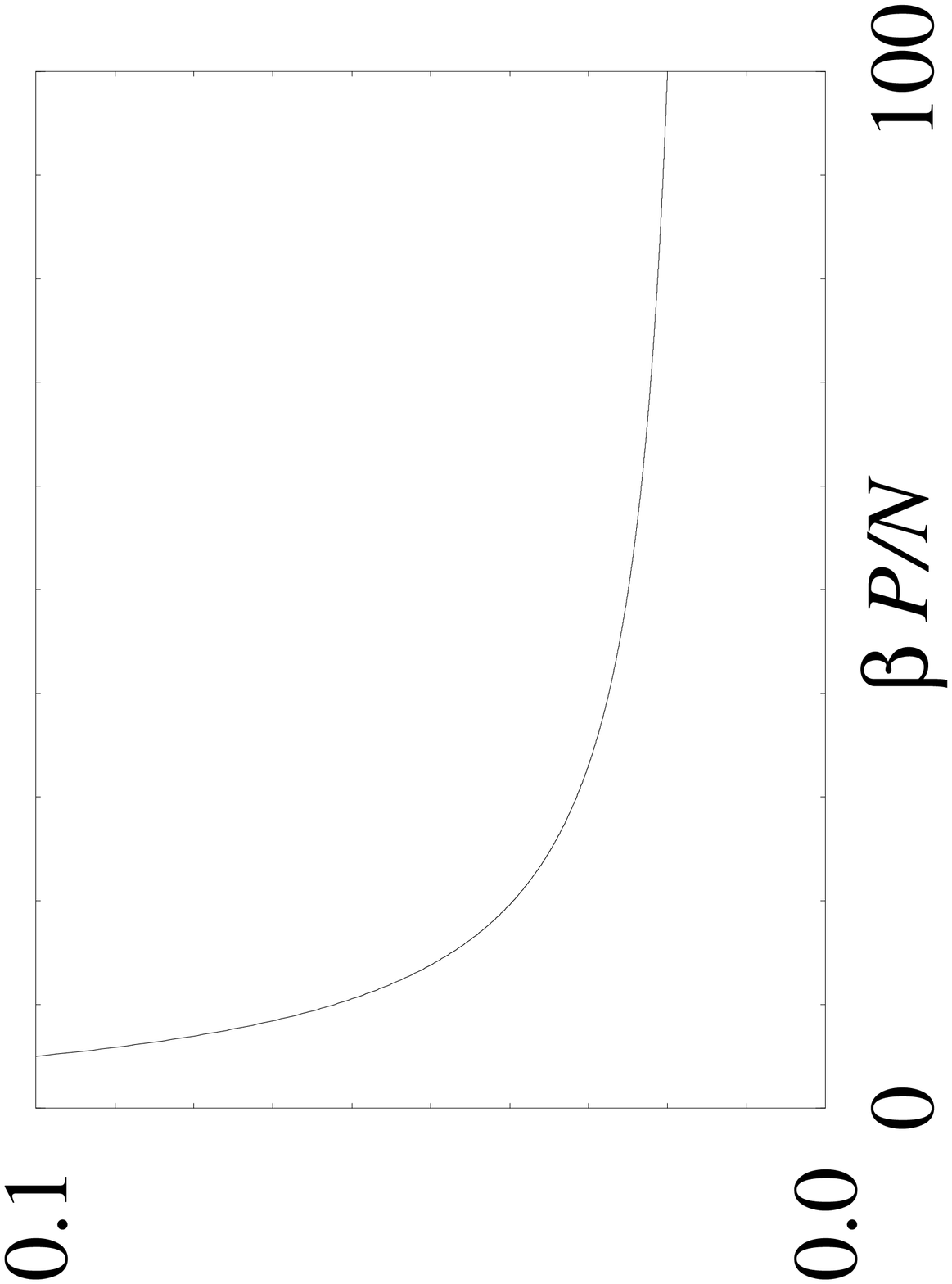}}}
\end{picture}
\end{center}
\caption{a) Learning curves for small $K$. The system with $K=2$ hidden units
 undergoes a second order transtion at $\alpha_c^{(2)} \approx 23.7$ leading
 to the kink in $\varepsilon_g(\alpha)$ (marked by the circle).  
 For $K\geq 3$ the transition is first order. Here we display the case $K=5$,
 solid lines correspond to the globally stable solution, dashed lines denote 
 local minima, ($\alpha_s^{(5)} \approx 44.3$, $\alpha_c^{(5)} \approx 46.6$ and
 $\alpha_d^{(5)} \approx 62.8$).  \protect\newline 
 b) The learning curve in the limit $K\to\infty$. The specialized state is a
 local minimum (dashed) for $ \alpha \geq  \alpha_s^{\infty}  \approx 61.0 $  and becomes
 globally stable (solid) at $\alpha_c^{\infty} \approx 69.1$. The unspecialized  configuration
 remains locally stable up to $\alpha_d^{K} = 4 \pi K$ for large $K$. The inset
 displays the initial (unspecialized) decrease of $\varepsilon_g$ with $\beta P/N$.}
\label{figure1}
\end{figure}

  The picture is qualitatively different for all $K \geq 3$ where we observe a first order  
  transition.  Again, for small $\alpha$
  the equilibrium solution is unspecialized $(R=S)$.  Then, for $\alpha \geq \alpha_s^{(K)} $
  a locally stable solution with $\Delta > 0$ and significantly lower generalization error appears.
  It becomes the global minimum at a critical value $\alpha_c^{(K)}$ where the free energies
  of the two solutions coincide. Finally, for 
  $\alpha \geq \alpha_d^{(K)}$, the local minimum with zero specialization disappears.
  Figure 1 (a) shows the learning curve for  $K=5$  as an example.  
  A more detailed description of the dependence of order parameters on $\alpha$ will be given
  in a forthcoming publication.

The difference in behavior between networks with $K\geq 3$ and $K=2$ is due to  the higher
degree of symmetry in the latter case. The permutation symmetry of hidden units results 
in a free energy, Eqs.\ (\ref{freesym},\ref{gensym}),
which is invariant under exchange  of $R$ and $S$ only for $K=2$. 
It is interesting to note that also the analysis of on--line learning in soft--committees
has revealed  qualitatively different features for $K=2$ and $K \geq 3$, see the discussion
of the fixed point structure in \cite{BRW}.

  Continuous Monte Carlo simulations of the learning process 
  confirm our findings qualitatively and show that the basic
  features of the specialization process remain the same for relatively low temperatures.
  Figure 2 (a) displays the  density  of observed student teacher overlaps 
  for $K=2$ close to equilibrium
  in the specialized and unspecialized phase ($\alpha=15$ and $\alpha=40$ respectively). 
  In panel (b) the corresponding histograms are plotted for a network with 
  $K=4$ at $\alpha=30$ and $\alpha=60$.
  Note that the total weight of overlaps $S$ should be a factor $(K-1)$ larger 
  than the contribution of type $R$ when hidden unit symmetry holds. 
  This is confirmed very well and justifies the simplifying assumption 
  (\ref{sitesymmetry}).

\begin{figure}[t]
\begin{center}
\setlength{\unitlength}{0.8pt}
\begin{picture}(325,100)(0,0)
 \put(0,0){\makebox(325,100)
          {\includegraphics{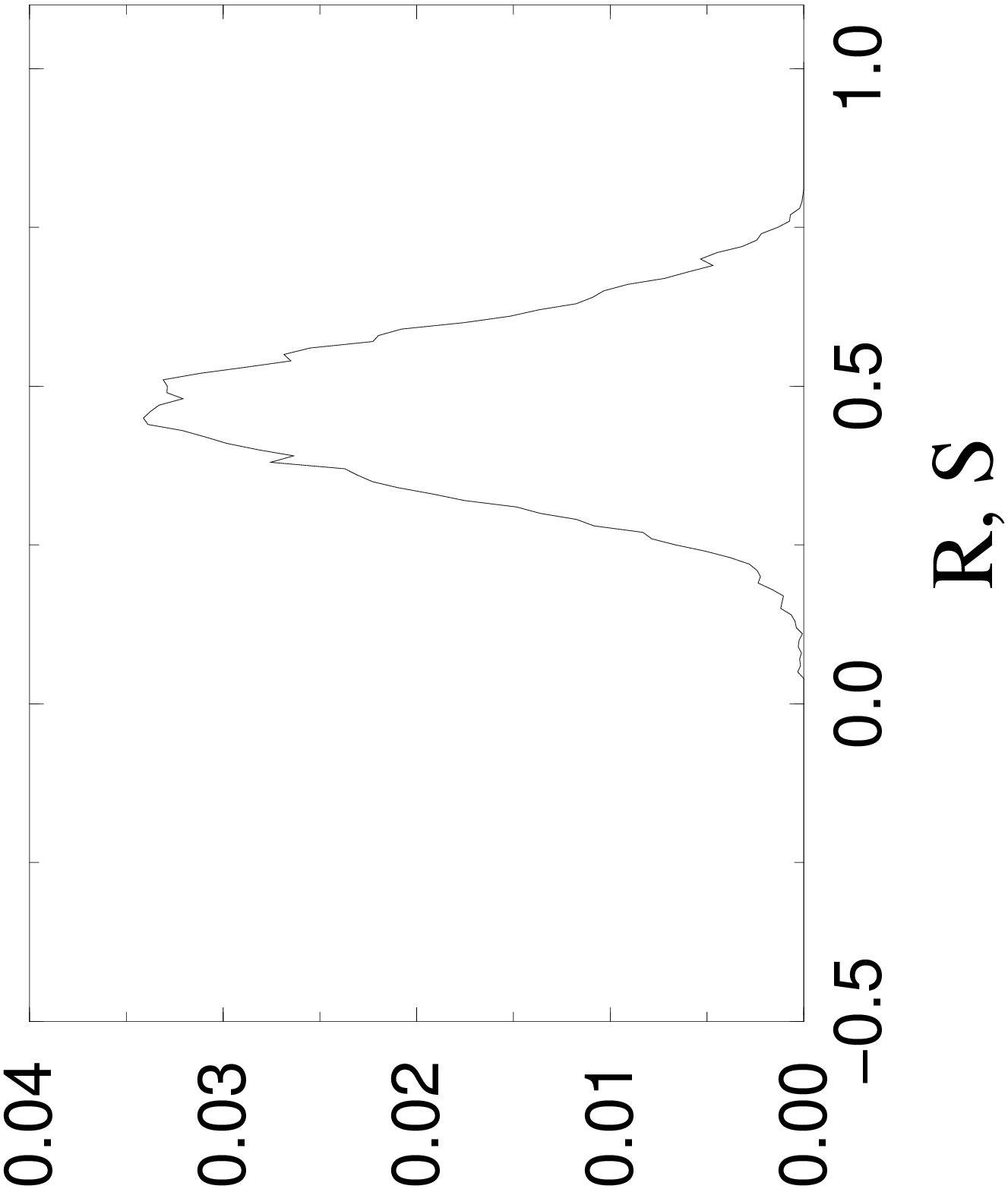}}}
 \put(0,0){\makebox(325,100)
          {\includegraphics{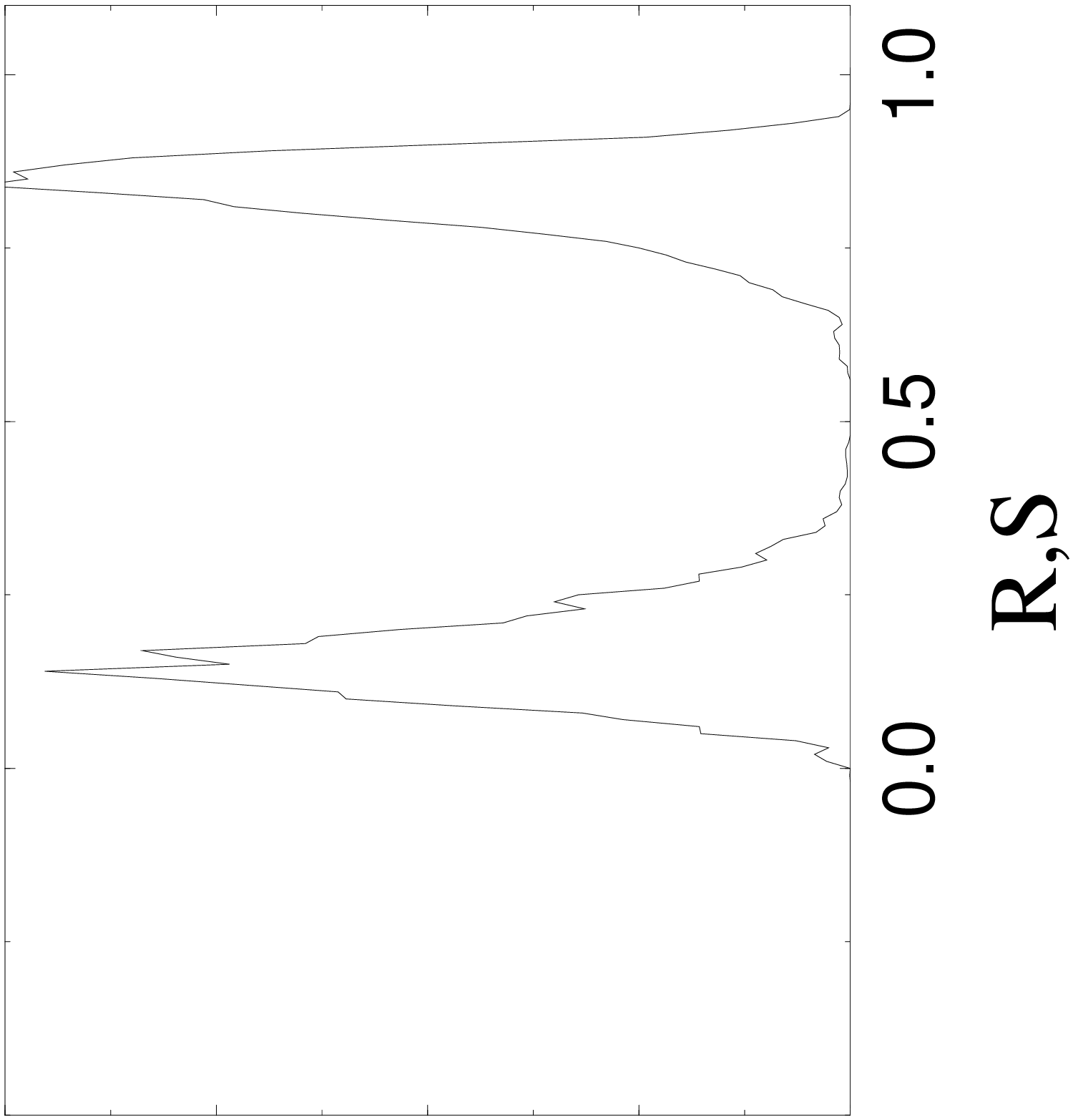}}}
  \put(0,0){\makebox(-140,-35){\bf a)}}
 \put(0,0){\makebox(325,100)
          {\includegraphics{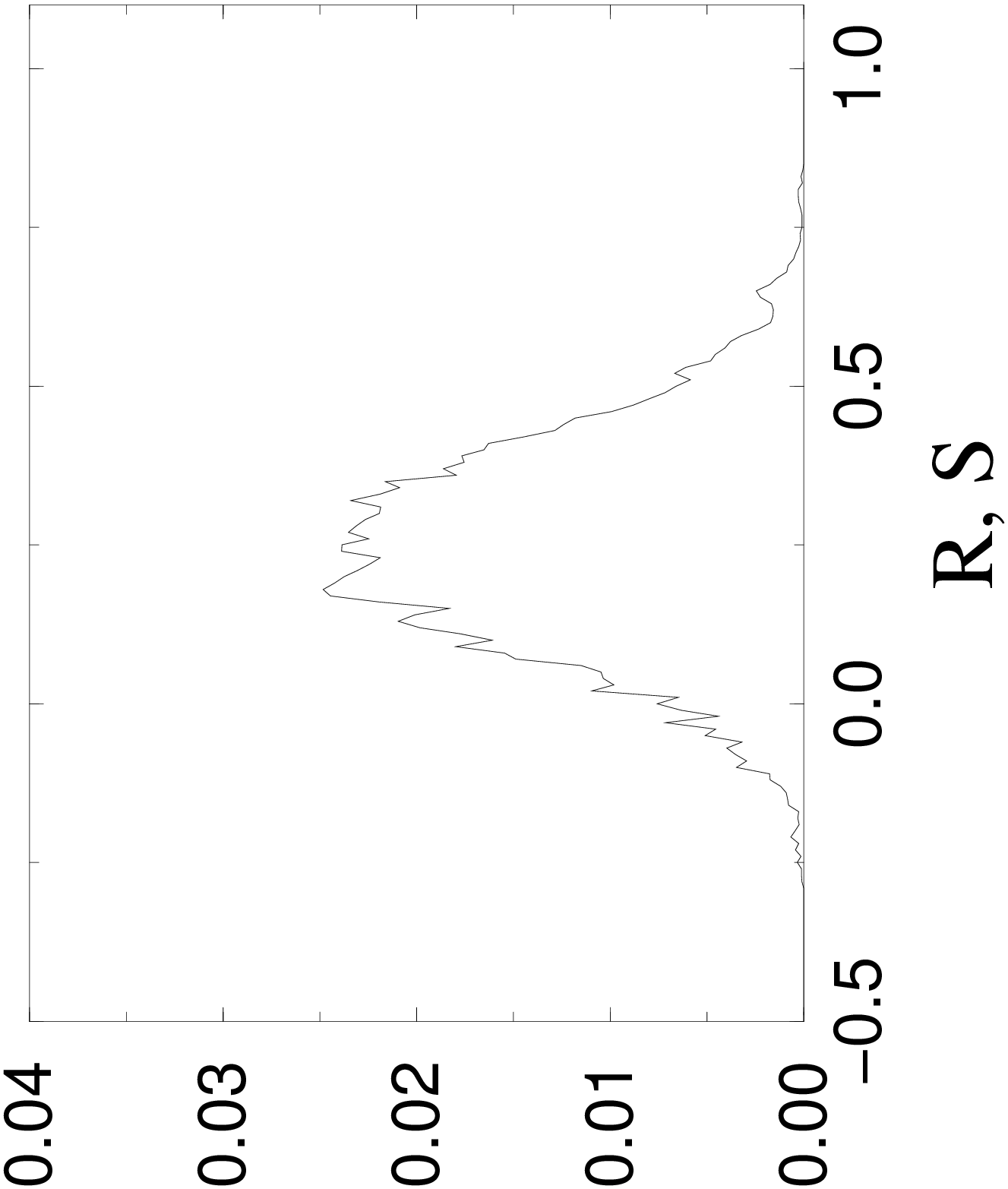}}}
 \put(0,0){\makebox(325,100)
          {\includegraphics{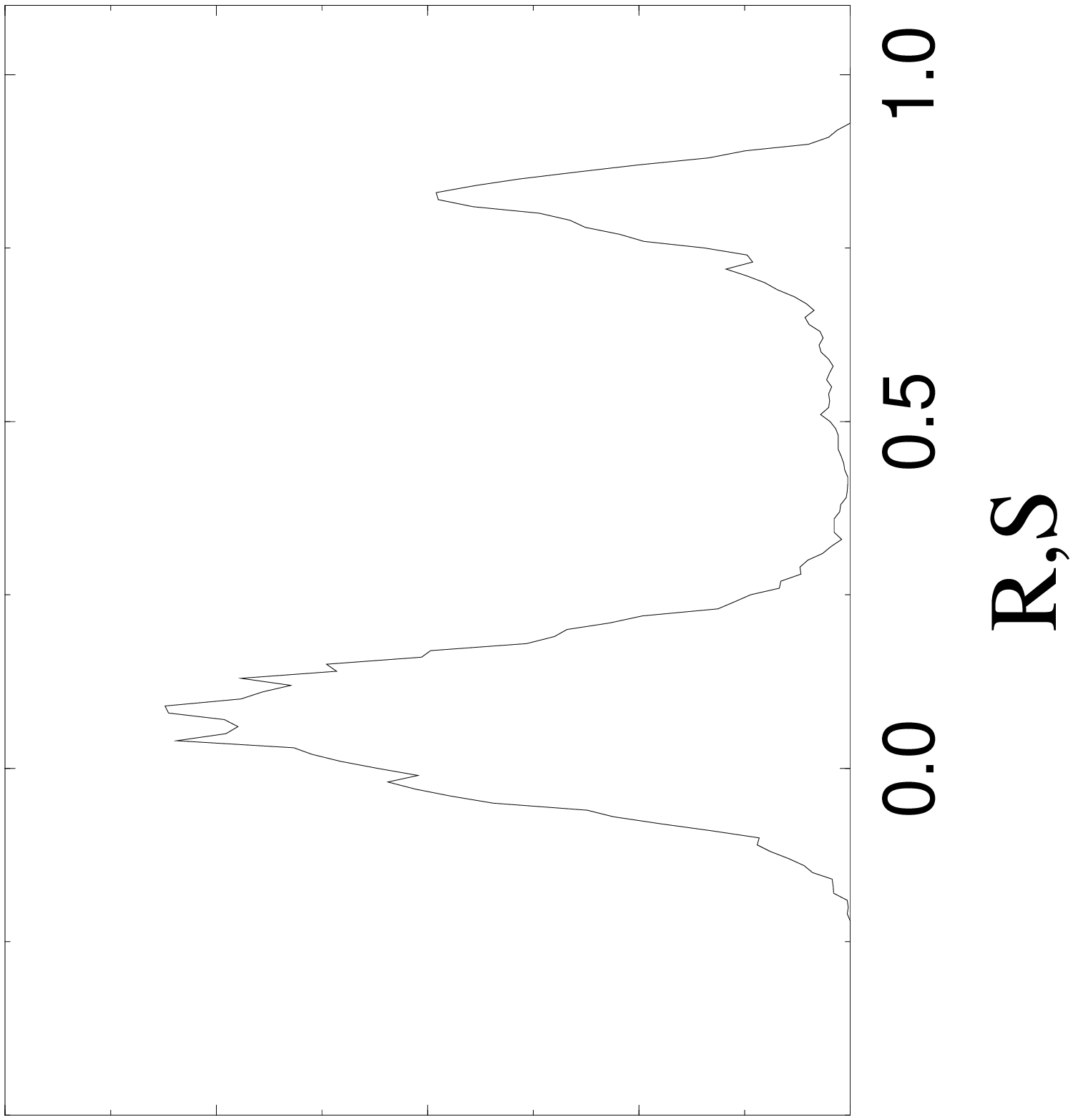}}}
 \put(0,0){\makebox(400,-35){\bf b)}}
\end{picture}
\end{center}
\caption{Empirical densities of observed student teacher overlaps in Monte Carlo
simulations  of the learning process at inverse temperature  $\beta = 0.8$. The system
size is $N=50$, after 50000 Monte Carlo steps allowed for equilibration, another
50000 were performed while sampling the densities.  
Panel a: $K=2$, $\alpha=15$ (left) and $\alpha=40$ (right),  
panel b: $K=4$, $\alpha=30$ (left) and $\alpha=60$ (right). 
   }
\label{figure2}
\end{figure}

 The behavior of very large networks 
 in the limit $K\to\infty$ (but $K <\!\!< N$)   
 has been studied in \cite{oh} for the transfer function $g(x)=\tanh(x)$
 within the annealed approximation. We repeat the discussion for $\beta \to 0$
 for completeness. Note that the analysis simplifies significantly due to 
 the choice $ g(x) = \erf(x/\sqrt{2})$.
 Different regimes have to be distinguished, in analogy to previous
 studies of large multilayered networks [14-18].
 %\cite{HolmJohn,Holmnur,oh,Robert}. 
 First, we assume that $ \beta P/N = \alpha K = {\cal O}(1) $
 and find that only  an unspecialized solution  exists in this initial phase of
 the learning process.  The ansatz  $ S = \widehat{S}/K $ and $ C = \widehat{C}/(K\!-\!1) $ yields a 
 simplified free energy      of the form 
  $  f =  (\alpha K) ( \widehat{C}/2 -\!\widehat{S} ) - \ln[1+\widehat{C}-\widehat{S}^2]/2 +
  \widehat{C}/2   + {\cal O}(1/K^2) $
 which is minimized  for 
 \begin{equation}  \textstyle
  \widehat{S} = \alpha K \left/ (\alpha K + \pi) \right.  \mbox{~~~and~~~} \widehat{C} = 
  -  \pi \alpha K \left/ ( \alpha K +\pi)^2 \right..
 \end{equation} 
 The corresponding generalization error is plotted in the inset of Figure 1 (b) {\it vs.\/}
 $\alpha K$ and approaches the stationary non--zero value  $ \varepsilon_g = 1/3 - 1/\pi $ 
 for large $\alpha K$.  

 Consequently, the unspecialized configuration is given by $ \widehat{S}= 1 $ and $ \widehat{C}=0 $ 
 to first order in $1/K$ when $\alpha = \beta P /(NK)$ is of order one.  
 However, specialization is possible in this
 regime and including a nonzero $ \Delta = R - S = {\cal O}(1) $ in the above ansatz
 yields  a solution  with  $ \widehat{S} = KS = 1-\Delta $ and $ \widehat{C}= KC= 0$  to first order. 
 The corresponding specialization $\Delta$ is the largest positive solution of  
 \begin{equation}
     \alpha \, = \frac{ \pi \Delta \sqrt{4-\Delta^2} }{ (1-\Delta^2) ( 2- \sqrt{4-\Delta^2} )}
 \end{equation}
 which does only exist for $ \alpha > \alpha_s^{\infty} \approx 60.99 $.  
The critical value $\alpha_c^{\infty} \approx 69.09$ is characterized by coinciding
 free  energies of the specialized and unspecialized solution.

 In the specialized phase, the generalization error decreases  asymptotically like
 $ \varepsilon_g  = 2/\alpha$  for $\alpha \to \infty$ which holds true also for 
 general (small) $K$. 
 The local minimum with  $\Delta=0$   remains locally stable   even in the limit
 $ \alpha \to \infty$.  However, we can show that
 the metastable state disappears at $ \alpha_d^{(K)} = 4 \pi K$ when K is large.  
 The latter result is an extremely good approximation for  $K$ as small as 4 or 5 
 already.  

  In summary, we have used equilibrium statistical physics to analyse off--line learning
   in two--layered soft--committee machines. First results for finite $K$  are obtained in the high
  temperature limit which allows to calculate the quenched free energy analytically.
  Specifically, we have studied networks with $K$ hidden units learning a perfectly matching
 rule. For $K=2$ we find a second order phase transition from unspecialized to  specialized 
 student configuration at a critical number of examples. In secenarios with $K\geq 3$ the
 transition from poor to good generalization is first order. Monte Carlo simulations confirm
 the existence and nature of the transitions also for moderately low temperatures.
  The analysis of the limit $K\to \infty$ shows that a critical number $ P \propto N K/\beta$
  of examples is needed for specialization, in agreement with the earlier findings 
  of Kang {et al.} \cite{oh}.  As a novel result we observe
  that the metastable unspecialized state  persists up to $ P = 4 \pi N K^2 / \beta $ for large $K$. 

  Our results (for $K\geq 3$) seem to parallel to a large extent the findings of [14-16]
  %\cite{HolmJohn,Holmnur,oh}
   for hard committee machines. We do not expect this correspondence to
  extend to very low temperatures, however. In the limit $T\to 0$, a transition to
  perfect generalization should occur after presenting on the order of ${\cal O}(NK)$ 
  examples, which is analogous to the results of \cite{boesetal} for a single continuous
  node.   This would not necessarily imply the existence of a corresponding practical
  algorithm. Note, however,  that already the computationally cheap  on--line gradient descent 
  realizes an exponential decay of $\epsilon_g$ with $\alpha = P/(NK)$ as opposed to the 
  much slower algebraic decay observed here. 

  It will therefore be necessary to complete the picture by extending the analysis into the
  low temperature regime, e.g.\ within the annealed approximation. The application of the 
  replica formalism should be possible for large networks in analogy to the work of
  \cite{HolmJohn,Robert}. 
  Further investigations will address unlearnable rules as well as over--sophisticated students
  \cite{Robert,SOK}.
  The introduction of a weight decay term to the training energy allows to relax the somewhat
  unnatural {\it a priori\/} normalization of student weights. First results concern a
  single unit and show a non--trivial dependence  of the performance on the weight
  decay parameter.  

  We would like to thank  W.\ Kinzel, G.\ Reents, and R.\ Urbanczik for  stimulating
  discussions, and  J.\ Hertz for drawing our attention to the analysis of 
  large soft--committees in \cite{oh}. \\


\begin{thebibliography}{99}
\bibitem{Hertz} J.A. Hertz, A.  Krogh  R.G. Palmer, 
  {\sl Introduction to the Theory of Neural Computation}, (Addison--Wesley, Redwood City, CA) 1991
\bibitem{KinzelOpper}  M. Opper and  W. Kinzel, in {\sl Models of Neural Networks III},
  eds. E. Domany, J.L. van Hemmen, and K. Schulten,  (Springer, Berlin) 1996
\bibitem{SST} S. Seung, H. Sompolinsky, and N. Tishby, Phys. Rev. A {\bf 45}, 6056, 1992
\bibitem{review}  T.L.H. Watkin, A. Rau, and M. Biehl,  Rev. Mod. Phys. {\bf 65}, 499, 1993
\bibitem{phase} W. Kinzel, Phil. Mag.  B  {\bf 77}, 1455, 1998
\bibitem{Cambridge} D. Saad (ed.), {\sl On--line learning in neural networks\/},
    Cambridge University Press, in press 
\bibitem{BiehlSchwarze}  M. Biehl and H. Schwarze,  J. Phys. A {\bf 28}, 643, 1995
\bibitem{SaadSolla}  D. Saad and S.A.  Solla, Phys. Rev. Lett {\bf 74}, 4337, 1995 and
 Phys. Rev. E {\bf 52}, 4225, 1995 
\bibitem{BRW}  M. Biehl, P. Riegler, and C. W\"ohler, J. Phys. A  {\bf 29}, 4769, 1996
\bibitem{local} R. Vicente and N. Caticha, J. Phys. A {\bf 30}, L559, 1997 
\bibitem{global} D. Saad and M. Rattray, Phys. Rev. Lett. {\bf 79}, 2578, 1997 
\bibitem{Chauvin} Y. Chauvin and D.E. Rumelhart (eds.), {\sl Backpropagation: Theory,
 Architecture, and Applications\/}, (Lawrence Erlbaum, Hillsdale, NJ) 1995
\bibitem{boesetal} S. B\"os, W. Kinzel, and M. Opper, Phys. Rev. E {\bf 47}, 1384, 1993
\bibitem{oh} K. Kang, J.--H. Oh, C. Kwon, and Y. Park, Phys. Rev. E {\bf 48}, 4805, 1993 
\bibitem{HolmJohn} H. Schwarze and J. Hertz, Europhys. Lett. {\bf 21}, 785, 1993 
\bibitem{Holmnur}  H. Schwarze, J. Phys. A {\bf 26}, 5781, 1993 
\bibitem{Robert} R. Urbanczik, J. Phys A {\bf 28}, 7097, 1995, and Phys. Rev. E, in press 
\bibitem{SOK} H. Schwarze, M. Opper, and W. Kinzel, Phys. Rev. A {\bf 46}, R6185, 1992
\end{thebibliography}
\end{document}